\documentclass{article}

\usepackage{arxiv}

\usepackage[utf8]{inputenc} 
\usepackage[T1]{fontenc}    
\usepackage{hyperref}       
\usepackage{url}            
\usepackage{booktabs}       
\usepackage{amsfonts}       
\usepackage{nicefrac}       
\usepackage{microtype}      
\usepackage{lipsum}
\usepackage{graphicx}
\graphicspath{ {./images/} }
\usepackage{amsmath}
\usepackage{amssymb} 
\usepackage{pifont}  

\newcommand{\xmark}{\ding{55}}
\usepackage{algorithm}
\usepackage{algpseudocode}
\usepackage{multirow}

\title{Efficient Long-Sequence Diffusion Modeling for Symbolic Music Generation}

\author{
    Jinhan Xu\textsuperscript{*} \\
    School of Computer Science and Artificial Intelligence \\
    Wuhan University of Technology \\
    Wuhan, China 430070 \\
    \texttt{375370@whut.edu.cn} \\
    \And
    Xing Tang\textsuperscript{\dag} \\
    School of Computer Science and Artificial Intelligence \\
    Wuhan University of Technology \\
    Wuhan, China 430070 \\
    \texttt{tangxing@whut.edu.cn} \\
    \And
    Houpeng Yang\textsuperscript{*} \\
    School of Computer Science and Artificial Intelligence \\
    Wuhan University of Technology \\
    Wuhan, China 430070 \\
    \texttt{houpengyang@whut.edu.cn} \\
    \And
    Haoran Zhang \\
    School of Computer Science and Artificial Intelligence \\
    Wuhan University of Technology \\
    Wuhan, China 430070 \\
    \texttt{339792@whut.edu.cn} \\
    \And
    Shenghua Yuan \\
    School of Computer Science and Artificial Intelligence \\
    Wuhan University of Technology \\
    Wuhan, China 430070 \\
    \texttt{yuanshenghua@whut.edu.cn} \\
    \And
    Jiatao Chen \\
    School of Computer Science and Artificial Intelligence \\
    Wuhan University of Technology \\
    Wuhan, China 430070 \\
    \texttt{346777@whut.edu.cn} \\
    \And
    Tianming Xie \\
    School of Computer Science and Artificial Intelligence \\
    Wuhan University of Technology \\
    Wuhan, China 430070 \\
    \texttt{tianming@whut.edu.cn} \\
    \And
    Jing Wang \\
    School of Computer Science \\
    Hubei University of Technology \\
    Wuhan, China 430068 \\
    \texttt{wangjing@hbut.edu.cn} \\
    \And
    Jiaojiao Yu\textsuperscript{\dag} \\
    School of Information Engineering \\
    Hubei University of Economics \\
    Wuhan, China 430205 \\
    Hubei Key Laboratory of Digital Finance Innovation \\
    Wuhan, China 430205 \\
    \texttt{jojo@hbue.edu.cn} \\
    \And
    Guangli Xiang \\
    School of Computer Science and Artificial Intelligence \\
    Wuhan University of Technology \\
    Wuhan, China 430070 \\
    \texttt{glxiang@whut.edu.cn}
    \thanks{Jinhan Xu and Houpeng Yang are co-first authors.}
    \thanks{Jiaojiao Yu and Xing Tang are the corresponding authors.}
}

\begin{document}
\maketitle
\begin{abstract}
Symbolic music generation is a challenging task in multimedia generation, involving long sequences with hierarchical temporal structures, long-range dependencies, and fine-grained local details. Though recent diffusion-based models produce high quality generations, they tend to suffer from high training and inference costs with long symbolic sequences due to iterative denoising and sequence-length-related costs. To deal with such problem, we put forth a diffusing strategy named SMDIM to combine efficient global structure construction and light local refinement. SMDIM uses structured state space models to capture long range musical context at near linear cost, and selectively refines local musical details via a hybrid refinement scheme. Experiments performed on a wide range of symbolic music datasets which encompass various Western classical music, popular music and traditional folk music show that the SMDIM model outperforms the other state-of-the-art approaches on both the generation quality and the computational efficiency, and it has robust generalization to underexplored musical styles. These results show that SMDIM offers a principled solution for long-sequence symbolic music generation, including associated attributes that accompany the sequences.
We provide a project webpage with audio examples and supplementary materials at \url{https://3328702107.github.io/smdim-music/}.
\end{abstract}

\section{Introduction}
{S}{ymbolic} music generation serves as a fundamental pillar of multimedia content production, entailing the synthesis and management of structured, time-evolving symbolic representations. \cite{huang2024symbolic,liang2024pianobart,tian2025xmusic}. Unlike low-level audio signals, symbolic music represents discrete musical events such as pitch, duration, rhythm, and temporal structure \cite{lerch2025survey}, which form a high-level abstraction of musical content that is widely used in composition, editing, and interactive creation \cite{bhandari2024motifs,luo2025bandcondinet}. Symbolic music as a long-sequence modality also has strong temporal dependencies and hierarchical structures along with long-range structural patterns \cite{ghoshal2026fusing}. These are important for maintaining musical coherence and expressiveness over long durations. Therefore, how to deal with the problems of long-term structure and fine-grained temporal information in symbolic music has become a core problem worthy of attention by the multimedia research community \cite{cheng2025fg}.

In symbolic music generation, abstractions such as a Musical Instrument Digital Interface sequence allow us to represent music and provide structure and thus ease of manipulation and interpretation. This is essential for automated music composition, recommendation, and analysis. \cite{huang2024symbolic,liang2024pianobart,kang2024video2music}, and captures higher-level musical attributes such as melodic motifs and thematic structures \cite{bhandari2025text2midi,li2025type}.

Diffusion-based generative models have shown strong performance in music generation \cite{mittal2021symbolic,zhang2025diffusion} and are well suited to symbolic music due to their robustness to discrete representations and capacity to model musical variability \cite{wang2024diffuseroll,zhang2024composer}. However, extending diffusion to long symbolic sequences introduces system-level challenges \cite{liu2024perturbing}: iterative denoising over high-dimensional token sequences leads to rapidly increasing computation and memory costs, further exacerbated by the need to model global dependencies at each step \cite{evans2024long,ou2025phrasevae}. 

To mitigate this issue, diffusion models augmented with Mamba-style state space models (SSMs) have been explored in the vision \cite{zhu2024vision}. Nevertheless, symbolic music consists of discrete event tokens with rich local dependencies that must be preserved throughout denoising, and directly transferring vision-oriented diffusion--SSM designs risks compromising token-level musical structure despite efficiency gains. As a result, effectively integrating SSM-based sequence modeling into diffusion for long-sequence symbolic music, while preserving both global coherence and fine-grained musical detail, remains an open problem.

In aggregate, current diffusion-based approaches are still inherently constrained when directly scaled up to long-sequence symbolic music generation \cite{wang2025via}. Most existing formulations simply assume we can execute each denoising step with the full sequence efficiently, an assumption that is quickly breaking as sequence lengths increase \cite{zhang2025inspiremusic}. Therefore, simply extending a diffusion pipeline to longer musical sequences is computationally prohibitive or yields poor global-structure modeling \cite{you2024momu}. Notably, this bottleneck remains unresolved by simple architectural substitutions, as it stems from a fundamental mismatch in nature between the diffusion--based denoising paradigm and the global--local requirements of long-form symbolic music \cite{naiman2024utilizing}. 

To deal with these problems, we put forward SMDIM, a novel framework for efficient long-sequence symbolic music generation based on diffusion modeling. SMDIM reformulates diffusion generation for extended temporal structures in order to better balance computational efficiency and the preservation of fine-grained musical detail. In contrast, to just optimizing for generation quality metrics, it directly targets the trade off between scale and expressiveness that arises in long-sequence diffusions. Thus, SMDIM allows coherent generation of longer musical sequences while requiring substantially less computational and memory resources.

Our contributions can be summarized as follows:
\begin{itemize}
    \item \textbf{A Unified Framework for Long-Sequence Diffusion:} 
    We propose SMDIM, a system-level framework tailored for long-sequence symbolic music generation under diffusion modeling. Instead of naively scaling existing diffusion pipelines, SMDIM is designed to explicitly address the balance between computational efficiency and generation quality that arises when applying diffusion models to temporally extended symbolic sequences.

    \item \textbf{Efficient Global--Local Modeling Mechanism:} 
    SMDIM uses a hybrid architecture that facilitates both efficient global sequence modeling and selectively preserve local musical details during the diffusion process. It allows the model to keep a long-range structural coherence without the prohibitive computational and memory cost of long-sequence diffusion without relying on a single architectural component.

    \item \textbf{Superior Performance Across Datasets:} 
    When evaluated on classical, pop, and traditional Chinese folk music datasets, including FolkDB (a previously introduced dataset for traditional Chinese folk music), SMDIM outperforms state-of-the-art methods in both generation quality and computational efficiency. FolkDB, which has been largely unexplored in prior work, further demonstrates SMDIM’s effectiveness in handling diverse musical styles, particularly those underrepresented in existing studies.

\end{itemize}

\section{Related Work}\label{sec:background}
\subsection{Symbolic Music Generation}
The generation of symbolic music has usually proceeded by treating it as either an image or as a sequence of discrete tokens. Image-based methods represent music as piano rolls, where pitch and time step are encoded to the space. These are processed with methods like CNNs or GANs \cite{wang2024style,cui2025music}. But these don’t have the ability to discern or extract temporal hierarchies of time within the music. Unlike this token-based approach views music as comprising discrete events. Such methods are popular for being able to properly model temporal structures \cite{wang2024whole,agarwal2024structure}. Following \cite{liang2024pianobart}, self attention based models have been shown to effectively capture long-range dependencies and generate coherent music sequences.

\subsection{Diffusion-Based Models for Symbolic Music Generation}
Diffusion models that were originally developed for images and audio generation also show promise in the area of music generation \cite{mittal2021symbolic,huang2024symbolic}. And these models use their capacity to capture the complex generative processes, iteratively refining the representations through steps of adding noise and denoising. In the context of symbolic music generation, diffusion models have also been applied to tasks such as melody composition and polyphonic music generation. However, the most existing diffusion-based symbolic music models rely on transformers style architectures and they exhibit limitations when applied to long sequences.

In regards to long sequences, most approaches primarily use Transformer architectures to model the sequential structure of music tokens \cite{plasser2023discrete,huang2024symbolic,min2023polyffusion}. These models can generate very coherent musical structures but suffer from a quadratic computational complexity due to the self-attention mechanisms they use. It is difficult to scale for longer sequences. Previous attempts combine existing long sequence generation techniques with diffusion models have been mostly unsuccessful. Non autoregressive approaches such as masked generative transformers \cite{chen2024musicldm} achieve efficiency at the cost of the iterative creative exploration of diffusion, often generating monotonous and conservative music. On the other hand, approaches such as hierarchical or staged diffusion force an artificial split between structure and detail generation.

\subsection{State Space Models and Diffusion for Symbolic Music Generation}
Recently state space models (SSMs) have received attention for long sequence modeling, as they are able to perform linear time inference and have a strong ability to model long term dependencies. Especially, Mamba \cite{gu2024mamba} has been successfully deployed in image generation, point cloud processing and time series forecasting \cite{zhu2024vision,zhang2024voxel,lambrechts2024parallelizing} with more efficient results than transformer based attention.

Several recent computer vision studies replace attention-heavy blocks in U-Net with bidirectional Mamba modules, which improves the efficiency of global context modeling during denoising \cite{teng2024dim}. This trend has also led to more investigation on combining SSM-based long-range sequence modeling with diffusion framework for long-form generative task, especially when quadratic attention becomes a bottleneck \cite{oshima2024ssm}.

These developments suggest the potential of diffusion--SSM integration for long-sequence generation. However, effectively adapting such designs to discrete symbolic music remains challenging, as global coherence and token-level musical details must be jointly preserved across denoising steps. This observation motivates our framework.More recently, PHALAR \cite{zhang2025mamba} addresses controllable symbolic music generation by representing music as image-like pianorolls rather than discrete tokens.

\section{Methods}\label{sec:experiment}
In this section, we present the methodological foundation and architectural design of our proposed model, SMDIM. Starting with the theoretical basis of SSMs and diffusion probabilistic models, we introduce the integration of these frameworks into a unified architecture tailored for symbolic music generation. Key innovations such as the MFA block and its hierarchical processing are discussed, highlighting how they balance efficiency and musical expressiveness.

\subsection{Problem Definition}

In symbolic music generation, the data consist of sequences of discrete tokens, such as pitch, duration, and timing attributes.
Let $\mathbf{x}_0 = (x_{0,1}, \ldots, x_{0,L})$ denote a clean symbolic music sequence of length $L$, where each token takes values from a finite vocabulary.
The forward diffusion process gradually corrupts $\mathbf{x}_0$ by stochastically replacing tokens with a special absorbing (mask) state over $T$ steps.
Specifically, the forward process is defined as a sequence of categorical transitions
\begin{equation}
q(\mathbf{x}_t \mid \mathbf{x}_{t-1})
= \prod_{i=1}^{L} q(x_{t,i} \mid x_{t-1,i}),
\end{equation}
where, at each step $t$, a token is either kept unchanged or replaced by the absorbing state according to a predefined noise schedule.
As $t$ increases, an increasing fraction of tokens are masked, and the process converges to a fully corrupted absorbing state $\mathbf{x}_T$.

The reverse process aims to iteratively recover the original sequence by learning a denoising model $p_\theta(\mathbf{x}_{t-1} \mid \mathbf{x}_t)$.
In practice, following standard masking-based discrete diffusion, the model is parameterized to directly predict the original clean tokens at masked positions, rather than predicting noise or transition probabilities.
This can be expressed in a factorized form as
\begin{equation}
p_\theta(\mathbf{x}_0 \mid \mathbf{x}_t)
= \prod_{i \in \mathcal{M}_t} p_\theta(x_{0,i} \mid \mathbf{x}_t),
\end{equation}
where $\mathcal{M}_t$ denotes the set of masked positions at diffusion step $t$. Starting from the absorbing state $\mathbf{x}_T$, the denoising model is applied
iteratively from $t=T$ to $t=1$, progressively reconstructing a coherent symbolic music sequence with long-range structure and fine-grained musical details.
From this perspective, symbolic music generation can be viewed as a multi-step sequence reconstruction problem under progressive masking, where the conditioning
on $\mathbf{x}_t$ is induced solely by the diffusion process itself and does not involve any external or semantic conditioning signals.

The formulation above follows a standard discrete diffusion objective with $\mathbf{x}_0$-prediction, without introducing additional supervision or task-specific loss terms.
The goal of this work is not to modify the diffusion objective itself, but to enable it to scale efficiently to long symbolic music sequences while maintaining
generation quality through framework-level design.

\subsection{Preliminary} \label{preliminary}
\subsubsection{Structured State Space Models}
SSMs are particularly suited for symbolic music generation due to their ability to efficiently capture long-range dependencies with linear complexity, making them scalable for long-sequence tasks. SSMs model the evolution of a system’s state through a state transition equation:
\begin{align}
x_t=Ax_{t-1}+\eta_t
\end{align}
where $x_t$ is the state vector, A is the transition matrix, and $\eta_t$ represents process noise, typically modeled as Gaussian noise. The observation equation relates the observed output $y_t$ (in our case, symbolic music tokens) to the current state $x_t$, as:
\begin{align}
y_t=Cx_t+\eta_t\end{align}
where $C$ is the observation matrix and $\eta_t$ represents observation noise. Typically modeled as $\eta_t \sim N(0,R)$.

We adopt Mamba, a structured state space model with selective state spaces, for its ability to model complex, high-dimensional data with low computational overhead. 
\subsubsection{Discrete Denoising Diffusion Probabilistic Models}
In this work, we choose our diffusion model to be a Discrete Denoising Diffusion Probabilistic Models(D3PM) \cite{austin2021structured} for symbolic music generation. D3PM extend diffusion models \cite{sohl2015deep} to discrete state spaces, where data is gradually corrupted over $T$ steps via a Markov chain. At each step, the forward process uses transition matrices $\{Q_t\}$ to update the data distribution:
\begin{align}
q(x_t | x_{t-1}) = \text{Cat}(x_t; p = x_{t-1} Q_t)
\end{align}
where $x_t$ is a one-hot vector and $Q_t$ is a row-stochastic matrix. The cumulative forward distribution over $T$ steps is:
\begin{align}
q(x_t | x_0) = \text{Cat}(x_t; p = x_0 Q_1 Q_2 \cdots Q_t)
\end{align}
To reverse this process, D3PMs parameterize the reverse transition probabilities $p_\theta(x_{t-1} | x_t)$, trained to approximate the true reverse process $q(x_{t-1} | x_t, x_0)$. The training objective minimizes the Evidence Lower Bound (ELBO), which consists of

\begin{itemize}
    \item 
      The Kullback-Leibler(KL) divergence between the forward distribution $q(x_T | x_0)$ and the prior $p(x_T)$,
    \item
      The KL divergence between forward and reverse transitions,
    \item
      The negative log-likelihood of reconstructing the original data $x_0$
\end{itemize}

\begin{figure}
    \centering
    \includegraphics[scale=0.050]{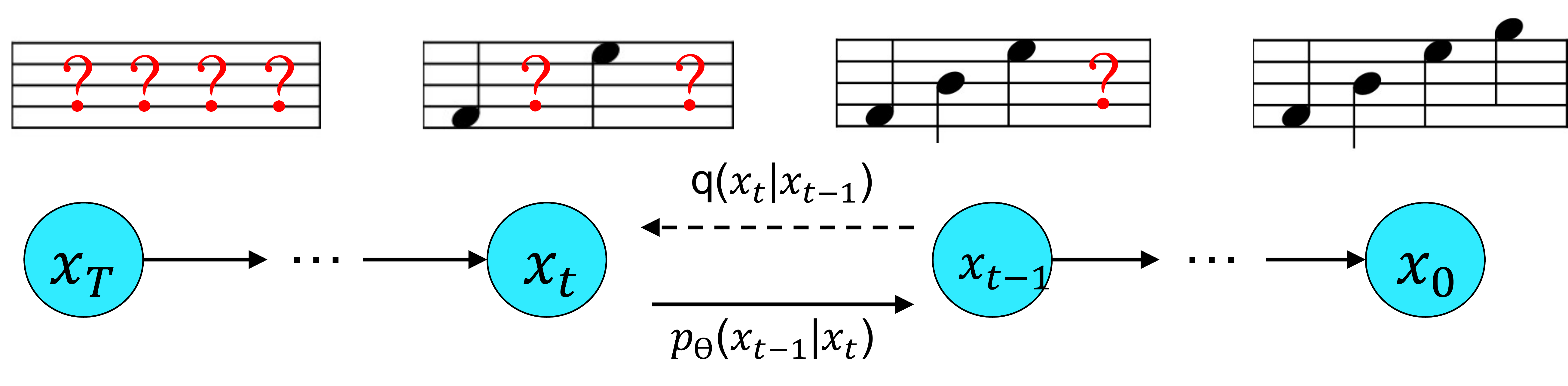}
    \caption{A score example illustrating the absorbing state diffusion process.}
    \label{fig:for_rev_diff}
\end{figure}

To simplify the forward process and ensure convergence, we adopt an \textbf{Absorbing State} (see Figure \ref{fig:for_rev_diff}) in the transition matrices $\{Q_t\}$:
\begin{align}
[Q_t]_{ij} = \begin{cases} 1, & \text{if } i = j = K \text{ (absorbing state)}, \\ 1 - \beta_t, & \text{if } i = j \text{ and } j \neq K, \\ \beta_t , & \text{if } j = K \text{ and } i \neq K, \end{cases}
\end{align}
where $\beta_t$ controls the corruption rate, and $K$ represents the absorbing state. This design ensures that as $t \to T$, all states converge to a deterministic distribution, enabling efficient computation and training on large-scale datasets \cite{sahoo2024simple}.

\subsection{Model Architecture}

\begin{figure}
    \centering
    \includegraphics[width=\textwidth]{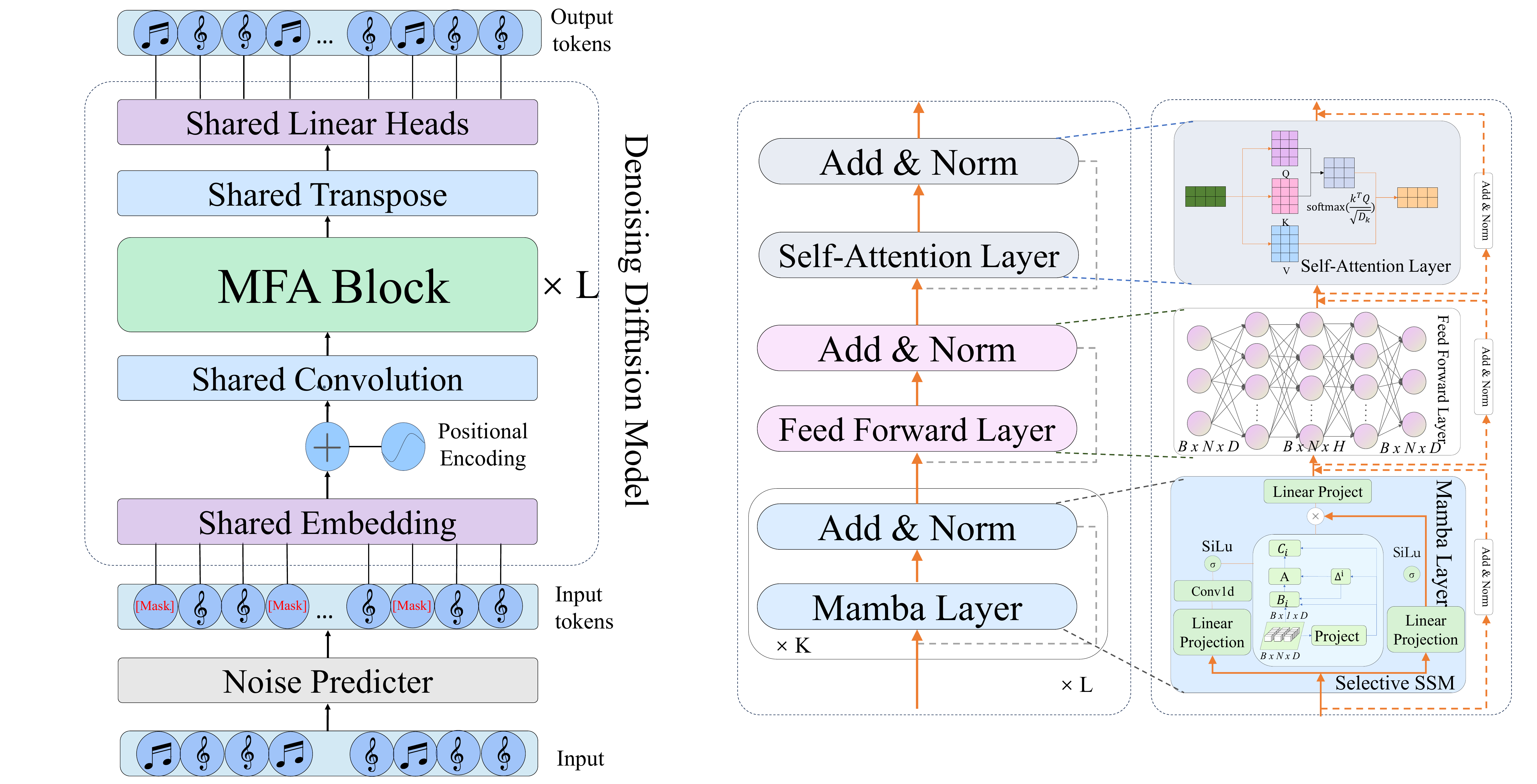} 
    \caption{The diagram shows SMDIM on the left and its core component, the MFA Block on the right. SMDIM processes input sequences hierarchically, while the MFA Block combines Mamba, FeedForward, and self-attention layers to balance scalability and precision. The [Mask] tokens represent noise, which is transformed into music symbols through denoising, resulting in coherent musical sequences.}
    \label{fig:SMDIM}
\end{figure}
The theoretical foundations discussed in Section \ref{preliminary} establish the basis for handling long-range dependencies (via SSMs) and modeling probabilistic sequences (via diffusion processes). However, directly applying these techniques to symbolic music generation presents unique challenges, particularly in balancing computational efficiency and the preservation of fine-grained musical details. 

To address these challenges, we propose a new architecture SMDIM,  which combines the advantages of state space models and diffusion frameworks. SMDIM is designed to handle long sequences without sacrificing musical coherence. SMDIM implements a hierarchical architecture to reconcile long-range structural dependencies with fine-grained local musicality. The following section details the SMDIM architecture and its constituent modules.

As depicted in Figure \ref{fig:SMDIM}, SMDIM employs a hierarchical structure to reconcile global structural coherence with fine-grained local details. First is a shared embedding layer that takes token indices and outputs continuous vectors. A shared 1D convolutional layer aggregates the embeddings into a shorter sequence with a richer feature representation, enabling the model to capture fine-grained local timing patterns. Then the compressed sequence is processed through multiple MFA blocks to extract context-aware representations. Finally, a shared transposed convolution layer reconstructs the sequence to its original resolution and a shared linear layer to output predictions. This design is effective for long-form symbolic music generation and results in improved structural coherence.

\subsection{MFA Block}



The MFA block is designed to simultaneously facilitate global context modeling and local detail preservation. To balance computational efficiency with the ability to model token-level details, we integrate self-attention layers, Mamba modules, and feed-forward networks within a unified architecture. This design addresses the respective limitations of state space models and transformer-based architectures.

MFA block is a design to deal with both long-range dependency and fine-grained musical details. Traditional state-space models can handle long sequences with linear-time complexity, but they may exhibit limitations in preserving token-level granularity due to the compression introduced by their sequence modeling mechanism. Self-attention preserves fine-grained details but scales quadratically with sequence length, making long-sequence modeling expensive. The MFA block mitigates this by combining efficient global modeling with selective local refinement.

Building on the hybrid design above, we adopt only one self-attention layer in each MFA block,
as a single attention layer is sufficient to recover the token-level local dependencies that may be lost 
during the SSM-based compression process. This design strikes a balance between representational 
capacity and computational efficiency. Similar behavior has also been observed in hybrid 
Attention--SSM architectures such as Jamba \cite{lenz2025jamba}, where sparse yet strategically placed attention layers 
provide the necessary inductive bias without requiring deeper attention stacks.

As illustrated in Figure \ref{fig:SMDIM}, the MFA block is composed of three key components:

\textbf{Mamba-layer} is designed to select relevant data and learn time-variant dependencies with a selective SSM. Note that $x$ is used as a shorthand $X_{\text{rep}}$, the process can be described as:
\begin{equation}
\begin{gathered}
x' = \sigma(\text{Conv1D}(\text{Linear}(x))), \\
z = \sigma(\text{Linear}(x)), \\
y' = \text{Linear}(\text{SelectiveSSM}(x') \odot z), \\
y = \text{LayerNorm}(y' + x)
\end{gathered}
\end{equation}
where $\sigma$ is the activation function SiLU, and $\odot$ denotes element-wise multiplication. The SelectiveSSM operation can be expressed as:
\begin{equation}
\begin{gathered}
\text{SelectiveSSM}(x') = y_t, \\
y_t = C h_t, \quad h_t = A h_{t-1} + B x'_t
\end{gathered}
\end{equation}
here, the matrices $A$, $B$ and $C$ are updated by a time-variant recurrent rule:
\begin{equation}
\begin{gathered}
B_t = S_B(x'_t), \quad C_t = S_C(x'_t), \\
\quad \Delta_t = \text{softplus}(S_A(x'_t))
\end{gathered}
\end{equation}
for the discrete parameters $A$ and $B$, we use:

\begin{equation}
\begin{gathered}
f_A(\Delta_t, A) = \exp(\Delta_t A) \\
f_B(\Delta_t, A, B_t) = (\Delta_t A)^{-1} (\exp(\Delta_t A) - I) \cdot \Delta_t B_t \\
A = f_A(\Delta_t, A), \quad B = f_B(\Delta_t, A, B_t)
\end{gathered}
\end{equation}
\textbf{FeedForward-layer}, inspired by the transformer architecture, introduces non-linearity :
\begin{equation}
\begin{gathered}
x_{\text{ffn}} = \text{FeedForward}(y_t; w_1, \sigma, w_2)
\end{gathered}
\end{equation}
where \(w_1\) and \(w_2\) are learnable  parameters, and \(\sigma\) is the activation function.\\
\textbf{Self-attention-layer} captures fine-grained token-level details while maintaining a global perspective across the sequence.
\begin{equation}
\begin{gathered}
x_{\text{att}} = \text{Softmax}\left(\frac{QK^\top}{\sqrt{d_k}}\right)VW^O,\\
 Q= x_{\text{ffn}}W_Q,K= x_{\text{ffn}}W_K,V= x_{\text{ffn}}W_V
\end{gathered}
\end{equation}
where \(W_Q\), \(W_K\), \(W_V\) are learnable parameters and scaled dot product is used.

\subsection{Efficiency Analysis}
In this section, we explore the computational efficiency of SMDIM by examining whether it is scalable for long-sequence symbolic music generation. The proposed architecture strategically combines Mamba, FeedForward, and self-attention modules to balance computational efficiency with modeling capacity.

\textbf{Computation Efficiency:}
We analysis the computational cost of our proposed MFA block, which includes Mamba, FeedForward and Self-attention layers. The key observation is that, aside from the self-attention component, the remaining modules achieve linear complexity with respect to the input sequence length $L$. The total computational complexity of the MFA block can be expressed as the sum of the complexities of the individual layers:

\begin{itemize}

\item \textbf{Mamba Layer:}
The Mamba layer introduces a quadratic dependency on the feature dimension $D$,
while remaining linear with respect to the sequence length $L$ and the state dimension $N$.
The expansion dimension is set to $E = 2D$ by default. Most of the computation cost is attributed to input projection, selective state-space modeling(SSM) and output projection.
Ignoring lower-order terms, the overall computational complexity of the Mamba layer is
\begin{align}
\Omega(\text{Mamba}) \approx 4LEN+3LED = 8LDN + 6LD^2 .
\end{align}

\item \textbf{FeedForward Layer:}
FeedForward Network (FFN) introduces a quadratic dependency on feature dimension $D$
while maintaining linear complexity with respect to the sequence length $L$.
The major cost comes from the 2 linear transforms from $D$ to the hidden dimension
\begin{align}
\Omega(\text{FFN}) = 8LD^2 .
\end{align}

\item \textbf{Self-Attention Layer:}
The self-attention mechanism suffers from a quadratic dependency of $L$ due to pairwise interactions between all tokens. Although linear projections have a cost of $O(LD^2)$, it’s mainly dominated by parameter generation, computing the attention matrix, and weighted aggregation.
The total computational complexity is expressed as:
\begin{align}
\Omega(\text{Self-Attention}) = 2L^2D + 4LD^2.
\end{align}

\item \textbf{Overall MFA Complexity:}
An MFA block consists of multiple Mamba layers, one FFN layer, and one self-attention layer.
The total computational complexity is therefore given by
\begin{align}
\Omega(\text{MFA}) = 8LDN + 18LD^2 + 2L^2D .
\end{align}
SMDIM restricts self-attention to a single layer in each MFA block, which improves scalability while preserving token-level modeling capacity. This design makes SMDIM well suited for long-sequence symbolic music generation.
\item \textbf{Threshold Analysis:}
Since the computational cost of self-attention grows quadratically with $L$, whereas the Mamba and FFN components scale linearly with $L$,
self-attention becomes the dominant computational bottleneck when
\begin{align}
2L^2D > 8LDN + 18LD^2 .
\end{align}
Simplifying the expression, the critical sequence length can be derived as:
\begin{align}
 L_{C} = 9D + 4N .
\end{align}
When $L < L_C$, the linear terms dominate the computational cost; whereas when $L > L_C$, the self-attention term becomes the primary bottleneck .
\end{itemize}

\textbf{Comparison to transformer-based Models:}
Traditional transformers that use self-attention heavily have quadratic complexity. It is impractical for long sequences as those that would arise from high-res music generation. On the other hand, SMDIM SMDIM predominantly relies on Mamba blocks and FFN layers, allowing the overall computation to scale approximately linearly with sequence length while capturing both global structure and fine-grained local details. This design significantly reduces GPU memory consumption and accelerates both training and inference, enabling SMDIM to handle longer sequences more effectively than transformer-based diffusion models.

\subsection{Additional Method Comparison}\label{app:method_comparison}
Table~\ref{tab:method_comparison} presents the main differences between SMDIM and representative baselines with respect to the generative paradigm, and step-wise design decisions which are key for efficient long-sequence symbolic music generation. Both autoregressive methods such as MusicTr and GETMusic use transformers to capture global dependencies but do not reformulate the diffusion process. This makes them unable to incorporate the step-wise global--local modeling mechanism explicitly. SCHmUBERT follows a diffusion paradigm but does full-sequence denoising with heavy attention-based architectures, limiting its scalability to long symbolic sequences. MusicMamba improves long-range modeling efficiency with state space models but is based on an autoregressive formulation, not explicitly addressing the iterative denoising dynamics and selective recovery of local musical details.

In contrast, SMDIM reformulates diffusive generation specifically for long symbolic sequences by means of a step-wise global–local design. At each denoising step, SMDIM combines SSM-based global modeling with lightweight attention for selective local recovery, enabling efficient long-range dependency modeling while preserving fine-grained token-level musical structure. This combination of diffusive reformulation and step-wise global–local modeling is missing in existing baselines and is responsible for the increased scalability and generation quality we observe in our experiments.

\section{Experiments}\label{sec:results}

\begin{table}[t] 
\caption{Comparison of architectural and generative designs for long-sequence music generation. (Diff.: Diffusion; AR: Autoregressive; G--L: Global--Local)}
\label{tab:method_comparison}
\centering
\setlength{\tabcolsep}{3pt} 
\renewcommand{\arraystretch}{1.2}
\footnotesize 
\begin{tabular}{l p{1.6cm} c c c c c}
\toprule
\textbf{Method} & \textbf{Backbone} & \textbf{Paradigm} & \textbf{Rep.} & \textbf{\shortstack{Diff.\\Ref.}} & \textbf{\shortstack{G--L\\Des.}} & \textbf{\shortstack{Loc.\\Rec.}} \\
\midrule
MusicTr \cite{huang2018music} & Transformer & AR & MIDI & \xmark & \xmark & \xmark \\
GETMusic \cite{lv2023getmusic} & Transformer & AR/S2S & Score & \xmark & \xmark & \xmark \\
SCHmUBERT \cite{plasser2023discrete} & Transformer & Diff. & VAE & \xmark & \xmark & \xmark \\
MusicMamba \cite{chen2025musicmamba} & SSM (Mamba) & AR & REMI & \xmark & \xmark & \xmark \\
\midrule
\textbf{SMDIM (Ours)} & \textbf{Hybrid} & \textbf{Diff.(Ref)} & \textbf{Token} & \textbf{\checkmark} & \textbf{\checkmark} & \textbf{\checkmark} \\
\bottomrule
\end{tabular}
\end{table}

\begin{table*}[h]
\caption{Objective evaluation of unconditional generation. The OA for 7 music attributes and the average OA are reported. The highest and second highest OA excluding GT are bolded and underlined respectively.}
\centering
\setlength{\tabcolsep}{10pt} 
\renewcommand{\arraystretch}{1.6} 
\resizebox{\textwidth}{!} {
\begin{tabular}{cccccccccc}
\toprule
Dataset & Method & Used Pitch & IOI & Pitch Hist & Pitch Range & Velocity & Note Duration & Note Density & Avg \\ 
\midrule
\multirow{7}{*}{\centering Maestro} 
& GT          & $0.905 \pm 0.127$ & $0.902 \pm 0.025$ & $0.955 \pm 0.009$ & $0.909 \pm 0.007$ & $0.910 \pm 0.281$ & $0.896 \pm 0.016$ & $0.917 \pm 0.015$ & $0.913 \pm 0.009$ \\
& MusicTr     & $0.631 \pm 0.092$ & $\underline{0.836 \pm 0.033}$ & $\underline{0.868 \pm 0.054}$ & $0.813 \pm 0.070$ & $0.604 \pm 0.048$ & $0.350 \pm 0.058$ & $\underline{0.894 \pm 0.012}$ & $0.755 \pm 0.016$ \\
& SCHmUBERT   & $0.429 \pm 0.059$ & $0.502 \pm 0.051$ & $0.701 \pm 0.030$ & $0.678 \pm 0.053$ & $0.417 \pm 0.074$ & $0.885 \pm 0.015$ & $0.773 \pm 0.046$ & $0.626 \pm 0.027$ \\
& Polyffusion & $\mathbf{0.803 \pm 0.020}$ & $0.831 \pm 0.024$ & $0.770 \pm 0.040$ & $0.814 \pm 0.017$ & $0.424 \pm 0.068$ & $0.786 \pm 0.021$ & $0.823 \pm 0.027$ & $0.750 \pm 0.013$ \\
& MusicMamba  & $0.551 \pm 0.027$ & $0.707 \pm 0.019$ & $0.669 \pm 0.012$ & $\underline{0.846 \pm 0.022}$ & $\mathbf{0.915 \pm 0.010}$ & $\mathbf{0.936 \pm 0.007}$ & $\mathbf{0.897 \pm 0.015}$ & $\underline{0.789 \pm 0.010}$ \\
& GETMusic    & $0.542 \pm 0.100$ & $0.799 \pm 0.045$ & $\mathbf{0.949 \pm 0.020}$ & $0.683 \pm 0.049$ & $\underline{0.883 \pm 0.026}$ & $0.857 \pm 0.032$ & $0.539 \pm 0.059$ & $0.750 \pm 0.016$ \\
& Ours        & $\underline{0.793 \pm 0.049}$ & $\mathbf{0.883 \pm 0.023}$ & $0.760 \pm 0.026$ & $\mathbf{0.866 \pm 0.034}$ & $0.871 \pm 0.026$ & $\underline{0.891 \pm 0.017}$ & $0.884 \pm 0.027$ & $\mathbf{0.850 \pm 0.018}$ \\
\midrule
\multirow{7}{*}{\centering Pop} 
& GT          & $0.881 \pm 0.026$ & $0.901 \pm 0.051$ & $0.957 \pm 0.011$ & $0.869 \pm 0.021$ & $0.911 \pm 0.026$ & $0.923 \pm 0.014$ & $0.903 \pm 0.023$ & $0.906 \pm 0.010$ \\
& MusicTr     & $\underline{0.807 \pm 0.041}$ & $\mathbf{0.904 \pm 0.026}$ & $0.810 \pm 0.036$ & $\underline{0.833 \pm 0.019}$ & $0.725 \pm 0.044$ & $0.735 \pm 0.068$ & $\mathbf{0.872 \pm 0.025}$ & $0.596 \pm 0.018$ \\
& SCHmUBERT   & $0.754 \pm 0.042$ & $0.236 \pm 0.031$ & $\underline{0.940 \pm 0.013}$ & $0.491 \pm 0.055$ & $0.808 \pm 0.032$ & $0.819 \pm 0.038$ & $0.465 \pm 0.048$ & $\underline{0.645 \pm 0.018}$ \\
& Polyffusion & $0.177 \pm 0.006$ & $\underline{0.888 \pm 0.036}$ & $0.544 \pm 0.030$ & $0.166 \pm 0.001$ & $0.804 \pm 0.033$ & $0.816 \pm 0.018$ & $0.756 \pm 0.061$ & $0.544 \pm 0.015$ \\
& MusicMamba  & $0.229 \pm 0.015$ & $0.151 \pm 0.012$ & $0.917 \pm 0.005$ & $0.156 \pm 0.013$ & $\underline{0.819 \pm 0.016}$ & $\mathbf{0.949 \pm 0.013}$ & $0.299 \pm 0.027$ & $0.503 \pm 0.007$ \\
& GETMusic    & $0.304 \pm 0.061$ & $0.404 \pm 0.024$ & $\mathbf{0.957 \pm 0.010}$ & $0.074 \pm 0.038$ & $0.715 \pm 0.059$ & $0.687 \pm 0.057$ & $0.520 \pm 0.041$ & $0.523 \pm 0.024$ \\
& Ours        & $\mathbf{0.881 \pm 0.018}$ & $0.862 \pm 0.033$ & $0.938 \pm 0.014$ & $\mathbf{0.906 \pm 0.014}$ & $\mathbf{0.828 \pm 0.033}$ & $\underline{0.914 \pm 0.020}$ & $\underline{0.868 \pm 0.025}$ & $\mathbf{0.885 \pm 0.012}$ \\
\midrule
\multirow{7}{*}{\centering Chinese Tradition} 
& GT          & $0.904 \pm 0.021$ & $0.881 \pm 0.016$ & $0.956 \pm 0.012$ & $0.914 \pm 0.012$ & $0.745 \pm 0.035$ & $0.910 \pm 0.014$ & $0.919 \pm 0.013$ & $0.890 \pm 0.006$ \\
& MusicTr     & $0.549 \pm 0.089$ & $\underline{0.808 \pm 0.031}$ & $0.854 \pm 0.037$ & $0.638 \pm 0.081$ & $0.651 \pm 0.038$ & $0.847 \pm 0.040$ & $0.844 \pm 0.029$ & $0.742 \pm 0.026$ \\
& SCHmUBERT   & $\mathbf{0.895 \pm 0.016}$ & $0.300 \pm 0.038$ & $0.882 \pm 0.018$ & $\mathbf{0.920 \pm 0.008}$ & $0.593 \pm 0.019$ & $0.362 \pm 0.028$ & $0.886 \pm 0.021$ & $0.691 \pm 0.010$ \\
& Polyffusion & $0.135 \pm 0.031$ & $0.791 \pm 0.038$ & $0.710 \pm 0.039$ & $0.338 \pm 0.059$ & $0.585 \pm 0.020$ & $0.781 \pm 0.024$ & $0.816 \pm 0.066$ & $0.594 \pm 0.013$ \\
& MusicMamba  & $0.796 \pm 0.060$ & $\mathbf{0.894 \pm 0.020}$ & $0.890 \pm 0.030$ & $0.837 \pm 0.055$ & $\underline{0.672 \pm 0.062}$ & $\underline{0.889 \pm 0.025}$ & $\mathbf{0.915 \pm 0.014}$ & $\underline{0.842 \pm 0.022}$ \\
& GETMusic    & $\underline{0.895 \pm 0.015}$ & $0.382 \pm 0.051$ & $\underline{0.950 \pm 0.018}$ & $\underline{0.917 \pm 0.014}$ & $0.299 \pm 0.042$ & $0.790 \pm 0.047$ & $0.724 \pm 0.053$ & $0.708 \pm 0.023$ \\
& Ours        & $0.891 \pm 0.016$ & $0.685 \pm 0.036$ & $\mathbf{0.962 \pm 0.007}$ & $0.901 \pm 0.011$ & $\mathbf{0.732 \pm 0.041}$ & $\mathbf{0.903 \pm 0.019}$ & $\underline{0.901 \pm 0.013}$ & $\mathbf{0.854 \pm 0.011}$ \\
\bottomrule
\end{tabular}}

\label{tab:model_comparison}
\end{table*}

\subsection{Data and Representation}

\textbf{Dataset:} We conduct experiments on a variety of different symbolic music datasets from different musical cultures and styles, such as MAESTRO \cite{hawthorne2018enabling}, POP909 \cite{wang2020pop909}, and FolkDB \cite{chen2025musicmamba}. MAESTRO consists of approximately 200 hours of Western classical piano performances, providing rich melodic and harmonic textures for analysis and modeling. POP909 dataset comprises about 60 hours of Chinese pop piano music which focuses on more contemporary harmonic language and repetition of structures. We additionally include FolkDB, an 11-hour dataset of traditional Chinese music that incorporates both pentatonic and heptatonic scales. This provides an evaluation of the model’s generalization to different modal systems and culturally specific rhythmic patterns.

\textbf{Data Representation:} We employ the REvamped MIDI-derived (REMI) representation \cite{huang2020pop} to encode MIDI data into discrete tokens that jointly represent local musical events and global structural information. In contrast to raw MIDI \cite{oore2020time}, REMI explicitly encodes note length, metrical structure, and harmonic information, enabling the model to learn rhythmic stability, phrasing, and harmonic structure. This structured representation enables SMDIM to handle long sequences while remaining musically expressive and structurally coherent across genres.

\textbf{Data Preprocessing:} Each musical piece is represented as a discrete REMI token sequence. The resulting sequence is treated as a single training sample $x_0 \in \mathcal{D}$, which is then used as clean data input for the diffusion.

\subsection{Training Process and Setup}
The training process of SMDIM follows a diffusion-based framework, as outlined in Algorithm 1, which provides a detailed description of the reverse diffusion training procedure. In particular, We consider the unconditional symbolic music generation setting, where the model learns a generative distribution over musical token sequences directly from data, without introducing explicit source--target pairs or conditional inputs. Table \ref{tab:self_Hyperparameters} summarizes the key hyperparameters and training configurations, which were carefully chosen to balance performance and computational feasibility. These settings allow the model to effectively capture complex musical patterns from large-scale symbolic datasets while maintaining manageable training times.


\begin{algorithm}[htbp]
\caption{Training Process for SMDIM}
\label{alg:training_process}
\begin{algorithmic}[1]
\State \textbf{Input:} Dataset $\mathcal{D}=\{\mathbf{x}_0^{(i)}\}_{i=1}^{N}$, where each $\mathbf{x}_0^{(i)}$ is a clean REMI token sequence; forward diffusion process $q(\mathbf{x}_t|\mathbf{x}_{t-1})$ with absorbing state; denoising model $p_\theta(\mathbf{x}_{t-1}|\mathbf{x}_t)$.
\State Initialize model parameters $\theta$.
\For{each training step $k = 1, \dots, K$}
    \State Sample a clean sequence $\mathbf{x}_0 \sim \mathcal{D}$.
    \State Sample a diffusion step $t \sim \mathcal{U}(\{1,\dots,T\})$.
    \State Sample corrupted sequence $\mathbf{x}_t \sim q(\mathbf{x}_t|\mathbf{x}_0)$ using the cumulative transition matrix $\bar Q_t$.
    \State Compute the analytic posterior $q(\mathbf{x}_{t-1}|\mathbf{x}_t,\mathbf{x}_0)$ from the forward process.
    \State Predict $p_\theta(\mathbf{x}_0|\mathbf{x}_t)$ using the denoising network.
    \State Construct $p_\theta(\mathbf{x}_{t-1}|\mathbf{x}_t)$ by combining $p_\theta(\mathbf{x}_0|\mathbf{x}_t)$ with $q(\mathbf{x}_{t-1}|\mathbf{x}_t,\mathbf{x}_0)$.
    \State Reweight ELBO by $w_t = (T - t + 1)/T$.
    \[
    \mathcal{L}_{\mathrm{vb}} =
    D_{\mathrm{KL}}\!\left(q(\mathbf{x}_T|\mathbf{x}_0)\,\|\,p(\mathbf{x}_T)\right)
    \]
    \[
    + \sum_{t=2}^{T} w_t \,
    \mathbb{E}_{q(\mathbf{x}_t|\mathbf{x}_0)} \!\left[
    D_{\mathrm{KL}}\!\left(
    q(\mathbf{x}_{t-1}|\mathbf{x}_t,\mathbf{x}_0)
    \,\|\, p_\theta(\mathbf{x}_{t-1}|\mathbf{x}_t)
    \right)\right]
    \]
    \[
    - \mathbb{E}_{q(\mathbf{x}_1|\mathbf{x}_0)}\!\left[\log p_\theta(\mathbf{x}_0|\mathbf{x}_1)\right].
    \]
    \State Update $\theta$ by minimizing $\mathcal{L}_{\mathrm{vb}}$ via gradient descent.
\EndFor
\Return Optimized parameters $\theta$.
\end{algorithmic}
\end{algorithm}

\begin{table}[h]
\caption{\label{tab:results:eval} Training Configuration and Hyperparameters.}
\centering
\setlength{\tabcolsep}{10pt} 
\scalebox{0.9}{
\begin{tabular}{ll}
\toprule
Component                   & Setting/Parameter          \\ 
\midrule
Framework                   & PyTorch                  \\
Hardware                    & 2 × NVIDIA RTX 4090        \\
Training duration           & 12 hours                   \\
Training steps              & 200,000                    \\
Optimizer                   & AdamW                      \\
Initial learning rate       & $5 \times 10^{-4}$         \\
Warm-up                     & First 10k steps            \\
Diffusion steps             & 1024                       \\
Batch size                  & 64                         \\
Input sequence length       & 2048                       \\
Generated music length      & 16 bars                    \\
Transformer embedding size  & 512                        \\
Transformer attention heads & 8                          \\
MFA block layers            & 8                          \\
LR schedule                 & Cosine annealing           \\ 
\bottomrule
\end{tabular}}
\label{tab:self_Hyperparameters}
\end{table}

\section{Evaluation}

\subsection{Baselines and Model Settings}
To benchmark SMDIM against representative state-of-the-art methods for symbolic music generation, we select five baseline models covering distinct architectural and training paradigms \cite{huang2018music,min2023polyffusion,plasser2023discrete,chen2025musicmamba,lv2023getmusic}. As summarized in Table~\ref{table:baseline}, all baselines are trained and evaluated on the same datasets, following the hyperparameters and optimization settings reported in the original publications whenever applicable.
The selected baselines employ different symbolic representations, reflecting that no single representation simultaneously optimizes expressivity, computational efficiency, and modeling convenience. Enforcing a unified representation would deviate from the original model designs and may misrepresent their reported performance. To ensure fair comparison, we use identical source data across models, together with representation-agnostic objective metrics and subjective listening tests, so that the relative quality and efficiency of SMDIM are assessed under the same dataset conditions.

\begin{table}[h]
\caption{Baselines for music generation.}
\centering
\setlength{\tabcolsep}{8pt} 
\scalebox{0.9} {
\begin{tabular}{lccc}
\toprule
\textbf{Method} & \textbf{Model} & \textbf{Dataset} & \textbf{Representation} \\ 
\midrule
MusicTr & transformer & Maestro & MIDI-like \\ 
PolyDiff & Diffusion & POP909 & Piano roll \\ 
SCHmUBERT & Diffusion & LMD & MusicVAE \\ 
GETMusic & Diffusion & Musescore & GetScore \\
MusicMamba & Mamba & FolkDB & REMI-M \\ 
\bottomrule
\end{tabular}}
\label{table:baseline}
\end{table}

\begin{table*}[h]
\caption{Ablation experiments for the design of SMDIM with various architectures. The order of layers significantly impacts the performance. The highest and second-highest OA excluding GT are bolded and underlined, respectively.}
\centering
\setlength{\tabcolsep}{10pt} 
\renewcommand{\arraystretch}{1.6} 
\resizebox{\textwidth}{!}{
\begin{tabular}{cccccccccc}
\toprule
Dataset & Method & Used Pitch & IOI & Pitch Hist & Pitch Range & Velocity & Note Duration & Note Density & Avg \\
\midrule
\multirow{7}{*}{\centering Maestro} 
& GT               & $0.905 \pm 0.127$ & $0.902 \pm 0.025$ & $0.955 \pm 0.009$ & $0.909 \pm 0.007$ & $0.910 \pm 0.281$ & $0.896 \pm 0.016$ & $0.917 \pm 0.015$ & $0.913 \pm 0.009$ \\
& Mamba Block      & $\underline{\mathbf{0.795 \pm 0.033}}$ & $0.151 \pm 0.016$ & $\mathbf{0.826 \pm 0.191}$ & $\underline{0.856 \pm 0.026}$ & $0.789 \pm 0.051$ & $0.413 \pm 0.073$ & $0.448 \pm 0.062$ & $0.605 \pm 0.033$ \\
& Transformer Block& $0.662 \pm 0.068$ & $\mathbf{0.911 \pm 0.017}$ & $0.762 \pm 0.030$ & $0.729 \pm 0.066$ & $\mathbf{0.906 \pm 0.021}$ & $0.888 \pm 0.018$ & $\underline{0.876 \pm 0.021}$ & $0.819 \pm 0.024$ \\
& AFM              & $0.792 \pm 0.049$ & $0.832 \pm 0.030$ & $\underline{0.826 \pm 0.027}$ & $0.851 \pm 0.045$ & $0.843 \pm 0.031$ & $0.820 \pm 0.026$ & $0.868 \pm 0.035$ & $\underline{0.833 \pm 0.020}$ \\
& FMA              & $0.724 \pm 0.076$ & $0.891 \pm 0.018$ & $0.737 \pm 0.029$ & $0.849 \pm 0.043$ & $0.848 \pm 0.034$ & $\underline{0.891 \pm 0.016}$ & $0.873 \pm 0.021$ & $0.830 \pm 0.021$ \\
& MFA (2SA)        & $0.805 \pm 0.048$ & $0.803 \pm 0.037$ & $0.812 \pm 0.028$ & $0.889 \pm 0.035$ & $0.751 \pm 0.046$ & $0.849 \pm 0.037$ & $0.843 \pm 0.025$ & $0.822 \pm 0.026$ \\
& Ours             & $\underline{0.793 \pm 0.049}$ & $0.883 \pm 0.023$ & $0.760 \pm 0.026$ & $\mathbf{0.866 \pm 0.034}$ & $\underline{0.871 \pm 0.026}$ & $\mathbf{0.891 \pm 0.017}$ & $\mathbf{0.884 \pm 0.027}$ & $\underline{\mathbf{0.850 \pm 0.018}}$ \\
\midrule
\multirow{7}{*}{\centering Pop} 
& GT               & $0.881 \pm 0.026$ & $0.901 \pm 0.051$ & $0.957 \pm 0.011$ & $0.869 \pm 0.021$ & $0.911 \pm 0.026$ & $0.923 \pm 0.014$ & $0.903 \pm 0.023$ & $0.906 \pm 0.010$ \\
& Mamba Block      & $\underline{0.852 \pm 0.030}$ & $0.179 \pm 0.034$ & $0.882 \pm 0.015$ & $0.784 \pm 0.047$ & $0.470 \pm 0.026$ & $0.872 \pm 0.036$ & $\underline{0.669 \pm 0.071}$ & $0.673 \pm 0.019$ \\
& Transformer Block& $0.302 \pm 0.051$ & $0.103 \pm 0.031$ & $0.547 \pm 0.026$ & $0.209 \pm 0.032$ & $0.179 \pm 0.005$ & $0.398 \pm 0.051$ & $0.112 \pm 0.042$ & $0.264 \pm 0.020$ \\
& AFM              & $0.837 \pm 0.030$ & $\underline{0.347 \pm 0.029}$ & $\underline{\mathbf{0.952 \pm 0.010}}$ & $0.833 \pm 0.026$ & $\underline{0.878 \pm 0.020}$ & $0.908 \pm 0.011$ & $0.590 \pm 0.052$ & $\underline{0.763 \pm 0.011}$ \\
& FMA              & $0.844 \pm 0.038$ & $0.011 \pm 0.006$ & $\underline{0.951 \pm 0.013}$ & $\underline{0.871 \pm 0.024}$ & $\underline{\mathbf{0.928 \pm 0.012}}$ & $\underline{0.911 \pm 0.027}$ & $0.176 \pm 0.045$ & $0.670 \pm 0.010$ \\
& MFA (2SA)        & $0.846 \pm 0.082$ & $0.008 \pm 0.001$ & $\underline{\mathbf{0.965 \pm 0.024}}$ & $0.877 \pm 0.021$ & $0.925 \pm 0.036$ & $\underline{\mathbf{0.926 \pm 0.092}}$ & $0.206 \pm 0.070$ & $0.679 \pm 0.025$ \\
& Ours             & $\underline{\mathbf{0.881 \pm 0.018}}$ & $\underline{\mathbf{0.862 \pm 0.033}}$ & $0.938 \pm 0.014$ & $\underline{\mathbf{0.906 \pm 0.014}}$ & $0.828 \pm 0.033$ & $0.914 \pm 0.020$ & $\underline{\mathbf{0.868 \pm 0.025}}$ & $\underline{\mathbf{0.885 \pm 0.012}}$ \\
\midrule
\multirow{7}{*}{\centering Chinese Tradition} 
& GT               & $0.904 \pm 0.021$ & $0.881 \pm 0.016$ & $0.956 \pm 0.012$ & $0.914 \pm 0.012$ & $0.745 \pm 0.035$ & $0.910 \pm 0.014$ & $0.919 \pm 0.013$ & $0.890 \pm 0.006$ \\
& Mamba Block      & $0.889 \pm 0.018$ & $0.522 \pm 0.033$ & $0.889 \pm 0.020$ & $\mathbf{0.916 \pm 0.015}$ & $\mathbf{0.755 \pm 0.033}$ & $0.886 \pm 0.012$ & $0.875 \pm 0.027$ & $0.819 \pm 0.010$ \\
& Transformer Block& $0.881 \pm 0.019$ & $0.570 \pm 0.048$ & $0.933 \pm 0.012$ & $\underline{0.914 \pm 0.009}$ & $0.673 \pm 0.020$ & $\underline{0.901 \pm 0.014}$ & $0.878 \pm 0.016$ & $0.821 \pm 0.012$ \\
& AFM              & $\underline{0.892 \pm 0.023}$ & $0.655 \pm 0.043$ & $\underline{0.957 \pm 0.007}$ & $0.903 \pm 0.020$ & $0.698 \pm 0.017$ & $0.892 \pm 0.013$ & $\underline{0.899 \pm 0.019}$ & $0.842 \pm 0.010$ \\
& FMA              & $\mathbf{0.893 \pm 0.019}$ & $\mathbf{0.714 \pm 0.049}$ & $0.948 \pm 0.013$ & $0.900 \pm 0.014$ & $0.699 \pm 0.028$ & $0.885 \pm 0.026$ & $0.898 \pm 0.017$ & $0.848 \pm 0.006$ \\
& MFA (2SA)        & $0.877 \pm 0.018$ & $0.748 \pm 0.045$ & $0.948 \pm 0.010$ & $0.914 \pm 0.013$ & $0.706 \pm 0.020$ & $0.882 \pm 0.020$ & $0.895 \pm 0.023$ & $\underline{0.853 \pm 0.007}$ \\
& Ours             & $0.891 \pm 0.016$ & $0.685 \pm 0.036$ & $\mathbf{0.962 \pm 0.007}$ & $0.901 \pm 0.011$ & $\underline{0.732 \pm 0.041}$ & $\mathbf{0.903 \pm 0.019}$ & $\mathbf{0.901 \pm 0.013}$ & $\mathbf{0.854 \pm 0.011}$ \\
\bottomrule
\end{tabular}}

\label{tab:ablation_studies}
\end{table*}

\subsection{Objective evaluations}
\textbf{Metrics:}
Objective evaluation in symbolic music generation remains an open problem \cite{lerch2025survey}. We adopt the metric of the average overlap area (OA), as proposed by \cite{yang2020evaluation}. The OA metric quantifies the similarity between the distributions of specific musical attributes in the generated music and the original dataset by measuring the overlapping area between their intra-set and inter-set distributions. We evaluate seven key musical attributes (pitch range, note density, etc.). 

\textbf{Results:}
The objective evaluation results are presented in Table \ref{tab:model_comparison}, which shows the OA scores for each musical attribute across the three datasets: MAESTRO, POP909, and FolkDB. In the table, bolded values indicate the highest OA scores, while underlined values represent the second-highest scores.
The proposed SMDIM consistently achieves the highest mean OA across all evaluated datasets, which means it is much closer to the ground truth. While each baseline performs well on its respective dataset, SMDIM consistently achieves stronger overall quality and better scalability under long-sequence settings.

\begin{figure} 
  \centering
  \includegraphics[width=0.6\linewidth]{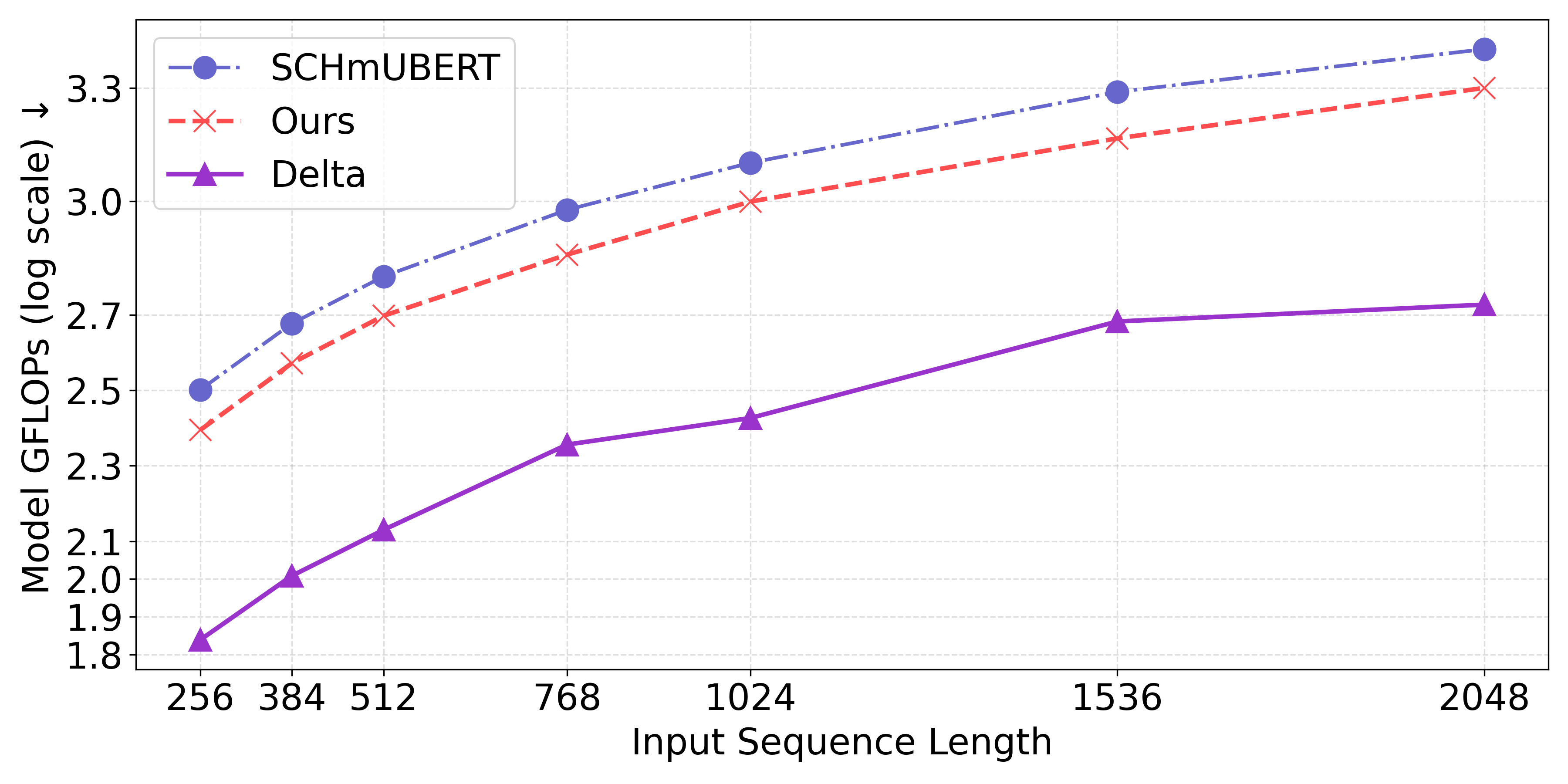} 
  \caption{Ablation results on GFLOPs of SMDIM model vs. SCHmUBERT model at different input seq. lengths. All GFLOPs are calculated by thop package for comparison.}
  \label{fig:gflops}
\end{figure}

\subsection{Efficiency Evaluations}
To validate the computational advantages of SMDIM, we compare its memory usage, execution time, and computational cost against SCHmUBERT \cite{plasser2023discrete}, a transformer-based diffusion model that also represents MIDI files as tokens. As shown in Table \ref{tab:efficiency_comparison}, SMDIM reduces memory consumption to 21 GB (vs. 35 GB for SCHmUBERT) and achieves a faster step time of 0.35 seconds (vs. 0.54 seconds), demonstrating its suitability for resource-constrained environments.
Figure \ref{fig:gflops} illustrates the efficiency of SMDIM in terms of GFLOPs across various input sequence lengths. For sequences of length 2048, the amount of GFLOPs by roughly 30\% and has a growing efficiency gap compared to SCHmUBERT as sequence length increases. These results suggest that SMDIM is well suited for long-sequence symbolic music generation, as its structured design reduces computational overhead without relying heavily on computationally intensive components.

\begin{table}[h]
\caption{Comparison of per-step latency and GPU memory usage with SCHmUBERT and MFA (2SA) on 2048-token sequences. Our method achieves the lowest latency and memory footprint.}
\centering
\setlength{\tabcolsep}{8pt} 
\scalebox{0.9}{
\begin{tabular}{ccc}
\toprule 
Method & Step Time (s) & GPU Memory (GB) \\
\midrule 
SCHmUBERT  & 0.54          & 35              \\
MFA (2SA)  & 0.63           & 24              \\
Ours       & \textbf{0.35} & \textbf{21}     \\
\bottomrule 
\end{tabular}}

\label{tab:efficiency_comparison}
\end{table}

\subsection{Long-Sequence Generation Stability}
To further test the stability of SMDIM with respect to generation length, we observe the statistical quality of generated music with respect to increase in output sequence length. Taking the Chinese Tradition dataset as a representative example, we measure the average Overlapping Area (OA) across multiple musical attributes for samples generated at different sequence lengths.

\begin{figure}
    \centering
    \includegraphics[width=0.6\linewidth]{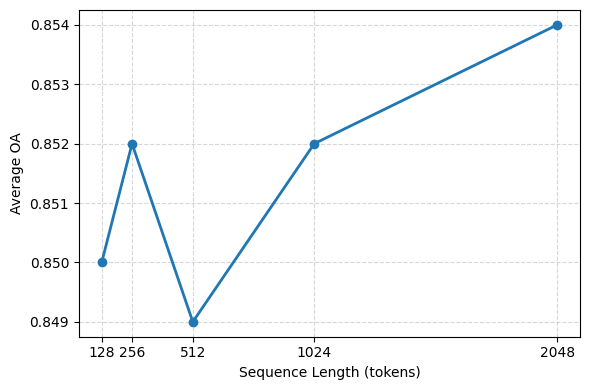}
    \caption{Average Overlapping Area (OA) of music generated by SMDIM under different output sequence lengths.}
    \label{fig:length_oa}
\end{figure}

As shown in Fig.~\ref{fig:length_oa}, the OA of SMDIM remains consistently stable as the generation length increases from 128 to 2048 tokens, with only minor fluctuations and no observable degradation at longer lengths, indicating robust generation quality under long-sequence settings.

\subsection{Ablation Studies}
To evaluate the performance of the Mamba layer, FeedForward layer, and Self-Attention layer within the MFA block, we conducted ablation studies. Table \ref{tab:ablation_studies} presents the results across the three datasets.

\begin{itemize}
    \item \textbf{Transformer Block:} 
    The transformer-only configuration achieves high scores on local-detail metrics such as IOI (e.g., 0.911 on MAESTRO). However, since it does not model the global effect, it obtains a lower score on pitch range and note density(e.g., 0.909, 0.896). The overall average score is less than the MFA block (0.833 < 0.850) which shows that using self-attention alone is limited.

    \item \textbf{Mamba Block:} 
    Using Mamba layers alone is good at global modeling metrics like Used Pitch and Pitch Range (e.g., 0.889, 0.916 on Chinese Tradition). However, the self-attention does not capture local token-level details, resulting in a low IOI and Velocity score (e.g., 0.832, 0.896 on MAESTRO). Self-attention is necessary for fine-grained representation.

    \item \textbf{AFM Reordering:} 
    Reorder MFA block operations as AFM: (Attention→FeedForward→Mamba) leads to consistent performance drop across all datasets. For example, the average OA score on MAESTRO decrease from 0.850 to 0.763, This implies that applying self-attention prior to Mamba-based compression may lead to the loss of global dependency information. In contrast, this result underscores the importance of the bottom-up refinement flow adopted in our design, where the Mamba layer first captures long-range dependencies, followed by non-linear refinement through the FeedForward layer, and finally local-detail enhancement via self-attention.
    \item \textbf{FMA Reordering:} 
    We also evaluate the FMA ordering to test whether moving the non-linear refinement from the end of the block to the beginning improves performance. The results indicate that FMA underperforms the original MFA design. When the FeedForward layer comes before Mamba, some of the early nonlinear projection is lost during the following state-space compression and so its effect is less in refining the tokens. Although Mamba can still capture global structure, the disruption of the progressive refinement pipeline leads to a consistent drop in OA scores across all datasets. This verifies that putting the FeedForward layer between Mamba and self-attention is necessary for keeping the hierarchical information flow.
    
    \item \textbf{MFA with Two Attention Layers:} 
    To explore if increasing the depth of the attention component would improve the expressiveness of the MFA block, we add a Double-SA variant with each MFA block containing two consecutive self-attention layers instead of one. For all datasets, the Double-SA design does in fact decrease the mean OA, implying that deeper attention does not imply better ability to model symbolic-music. Table~\ref{tab:efficiency_comparison} also confirms this: the Double-SA configuration has a much higher step time and memory footprint, but does not improve OA. From the above, it is evident that the original single attention MFA was already effective at finding an excellent tradeoff between performance and efficiency.

\end{itemize}
    
These experiments show how each MFA block component works together. The Mamba layer is computationally efficient and captures global dependencies, whereas the self-attention layer is more computationally expensive and focuses on token-level interactions. The FeedForward layer is used to refine the non-linear representations, which can integrate both global and local feature. The original processing order (Mamba → FeedForward → Attention) is essential for achieving the best performance across all datasets, as it balances global compression with the preservation of local details.

\begin{table}[h]
\caption{Demographic profiles of the 15 professional evaluators.}
\label{tab:demographic}
\centering
\setlength{\tabcolsep}{3.5pt} 
\renewcommand{\arraystretch}{1.1} 
\footnotesize 
\begin{tabular}{c c l l c l}
\toprule
\textbf{ID} & \textbf{Grp.} & \textbf{Profession / Major} & \textbf{Status} & \textbf{Yrs} & \textbf{Area of Expertise} \\
\midrule
P01 & Theo. & Composition & Asst. Prof. & 17 & Harmony, Counterpoint \\
P02 & Theo. & Music Theory & PhD Cand. & 14 & Form, Tonal Harmony \\
P03 & Theo. & Screen Scoring & Composer & 11 & Orchestration, Development \\
P04 & Theo. & Jazz Studies & Master Std. & 10 & Jazz Harmony, Improv. \\
P05 & Theo. & Comp. \& Arranging & Lecturer & 19 & Arrangement, Aural Skills \\
\midrule
P06 & Perf. & Piano Performance & Musician & 15 & Classical, Articulation \\
P07 & Perf. & Violin Performance & Master Std. & 12 & Ensemble, Intonation \\
P08 & Perf. & Accompaniment & Grad. Std. & 11 & Keyboard Harmony, Voicing \\
P09 & Perf. & Pop/Jazz Keyboard & Musician & 9 & Groove, Pop Progressions \\
P10 & Perf. & Conducting & Master Std. & 13 & Score Reading, Balance \\
\midrule
P11 & App. & Music Technology & Master Std. & 8 & MIDI Editing, Rendering \\
P12 & App. & Music Production & Producer & 12 & Mixing, Texture Control \\
P13 & App. & Musicology & PhD Cand. & 13 & Style Analysis, Genre \\
P14 & App. & Electronic Music & Indep. Musician & 10 & Sound Design, Synth \\
P15 & App. & Music Education & Grad. Std. & 7 & Pedagogy, Perception \\
\midrule
\textbf{Avg.} & -- & -- & -- & \textbf{12.7} & -- \\
\bottomrule
\end{tabular}
\end{table}

\subsection{Subjective Evaluations}
To ensure a fair subjective evaluation, a double-blind listening test protocol was adopted. The assessment involved 15 recruited participants—including university music professors, professional musicians, and graduate researchers—all of whom possess formal training in music composition, theory, or performance. The comprehensive demographic profiles of these professional evaluators are detailed in Table~\ref{tab:demographic}. For each dataset, each participant judged 20 randomly sampled music clips per model. The length of each audio clip was about 150 seconds, which was converted into waveform audio with fixed tempo and instruments from symbolic music. The scores for each method were averaged over all participants and evaluated samples.

The evaluation followed four commonly adopted dimensions in music perception research \cite{lerch2025survey}, which are closely related to established concepts in auditory and musical perception. Specifically, Emotional Consistency reflects listeners’ perception of affective coherence over time, Melodic Fluency corresponds to melodic continuity and pitch trajectory smoothness, Rhythmic Stability relates to temporal regularity and rhythm perception, and Structural Clarity captures the perceptual organization of long-term musical form and sectional coherence.

As shown in Table~\ref{tab:subjective}, our model achieved competitive or superior scores compared with all baseline methods, demonstrating its ability to generate musically coherent and stylistically consistent outputs, with particularly strong gains in Melodic Fluency and Structural Clarity. Statistical significance was assessed using paired two-sided t-tests, and the improvements in Melodic Fluency and Structural Clarity were found to be statistically significant (\(p < 0.05\)).

\begin{table}[h]
\caption{Subjective listening study results of the 4 evaluation aspects (1-10 scale).
Statistically significant improvements over the best baseline are marked with * (\(p < 0.05\), paired two-sided t-test).}
\centering
\setlength{\tabcolsep}{4pt}
\scalebox{0.9}{
\begin{tabular}{lcccc}
\toprule
\textbf{Models} 
& \textbf{Emotional} & \textbf{Melodic} & \textbf{Rhythmic} & \textbf{Structural} \\
& \textbf{Consistency} & \textbf{Fluency} & \textbf{Stability} & \textbf{Clarity} \\
\midrule
MusicTr    & 7.68 & 7.87 & 7.37 & 6.99 \\
SCHmUBERT  & \textbf{7.72} & 7.85 & \textbf{7.38} & 7.03 \\
PolyFusion & 7.63 & 7.83 & 7.34 & 6.98 \\
MusicMamba & 7.69 & 7.85 & 7.29 & 7.00 \\
Ours       & 7.66 & \textbf{7.90}$^{*}$ & 7.30 & \textbf{7.09}$^{*}$ \\
\bottomrule
\end{tabular}}
\label{tab:subjective}
\end{table}

\begin{figure}
  \centering
  \includegraphics[width=0.48\textwidth]{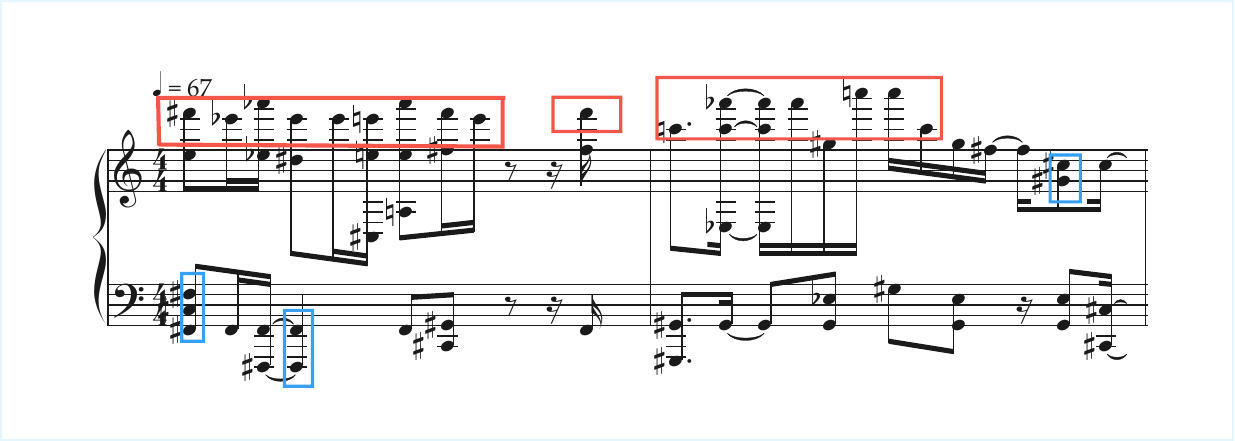}
  \hspace{0.02\textwidth}
  \includegraphics[width=0.48\textwidth]{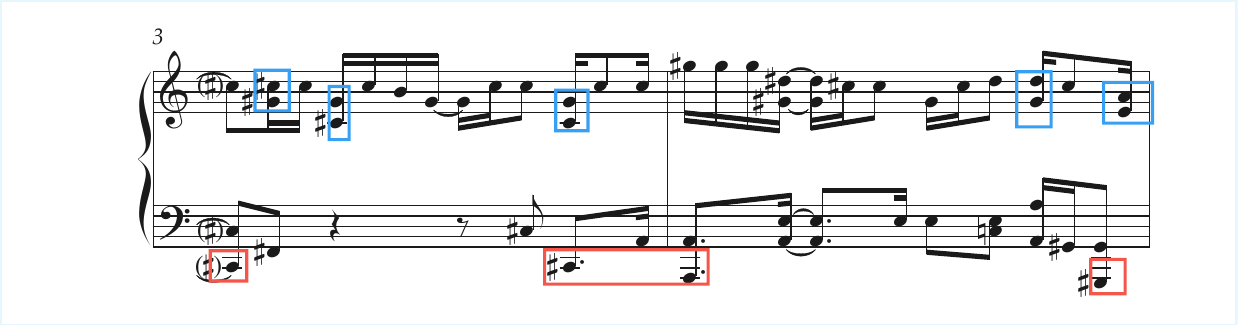}

  \vspace{-0.4em}

  \includegraphics[width=0.78\textwidth]{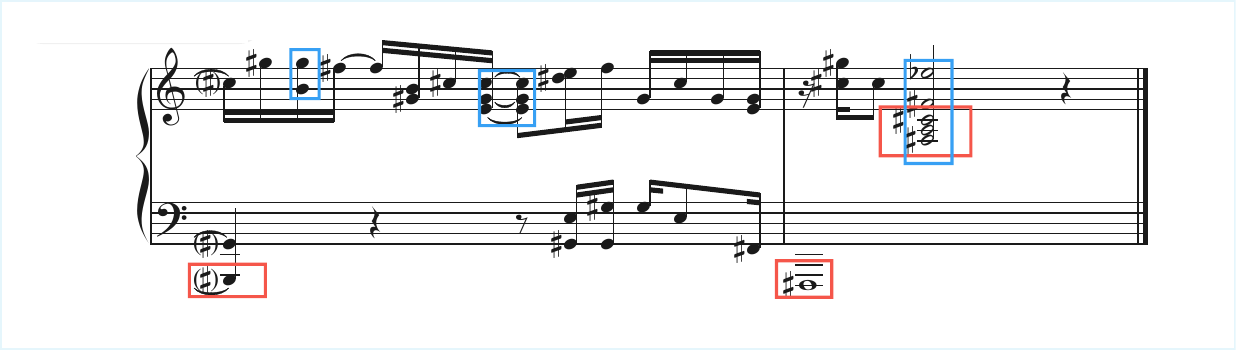}

  \vspace{-0.3em}

  \caption{Representative failure cases generated by the proposed model.
  (a)--(b) Failure cases exhibiting extreme pitch ranges and overly dense vertical note stacking,
  resulting in musically implausible symbolic structures.
  The regions highlighted in red and blue indicate representative local errors,
  including abnormal pitch outliers and excessive simultaneous note activations.
  (c) A failure case from the final portion of a generated sequence,
  illustrating structural degradation and weakened global musical organization over time.}
  \label{fig:failure_cases}
\end{figure}

\section{Failure Case Analysis}

Although the proposed model is good for both objective and subjective evaluations, we have some systematic failure cases under certain musical and structural conditions. These findings help reveal the remaining limitations of the model in symbolic music generation.

A representative failure mode involves the generation of notes with musically implausible pitch ranges. In such cases, the model generates pitch outliers that fall far outside the typical melodic range, leading to excessive ledger lines and musically implausible pitch contours. We attribute this behavior to the discrete and effectively unbounded pitch-token space in the symbolic representation: under high uncertainty, the generator may occasionally sample extreme values in the absence of explicit pitch-range regularization.

Another frequently observed issue is overly dense vertical note stacking. Instead of forming interpretable harmonic structures, the model sometimes produces several notes all at once at a single time step, creating an unnatural thickness of chordal texture. These observations suggest that the model lacks effective mechanisms to enforce sufficient harmonic sparsity and clear voice separation, particularly in rhythmically complex passages or in later stages of long generated sequences. In addition, structural coherence tends to degrade toward the end of longer compositions. Although local melodic continuity is generally preserved, global musical form and thematic coherence may gradually deteriorate over time. Such late-stage structural drift highlights the difficulty of maintaining long-term musical organization in diffusion-based symbolic generation, where small prediction errors can accumulate over extended denoising steps and gradually degrade global structure.

An example of some representative failure cases is shown in Fig.~\ref{fig:failure_cases}. Specifically, Fig.~\ref{fig:failure_cases}(a) and Fig.~\ref{fig:failure_cases}(b) represent extreme pitch range failures as well as over-dense vertical note stacking failures.
In those cases, it produces extremely high or low pitches and very thick chordal textures, resulting in musically implausible symbolic structures. The regions highlighted in red and blue illustrate representative local errors, including abnormal pitch outliers and excessive simultaneous note activations.
In contrast, the failure case in Fig.~\ref{fig:failure_cases}(c) exhibits a degradation of overall musical structure toward the end of a long generated sequence. This late-stage structural drift leads to reduced thematic coherence and weakened long-term organization, highlighting the challenge of maintaining global musical form over extended generation.

Accordingly, future work may explore incorporating explicit constraints on pitch range, harmonic sparsity, long-term musical structure, and higher-level musical priors into symbolic music generation models.

\section{Conclusion}

   In this paper, we introduce SMDIM, a diffusion-based architecture for Symbolic Music Generation that integrate Mamba, Feedforward and self-attention layers within the MFA block. Through experimental evaluations on different datasets like MAESTRO, POP909, and Chinese FolkDB, it shows its great performance and potential for long-sequence modeling and efficient music generation. SMDIM is effective at modeling both long-range dependencies and fine-grained local details; however, performance may degrade for highly complex musical structures and extremely long compositions. Future work will focus on improving its robustness to such complexities and further enhancing scalability.

\section*{Acknowledgments}

This work was supported in
part by the National Natural Science Foundation of China under Grant
62272356, Grant 62302155, and Grant 62372344.

\section*{Declaration of generative AI and AI-assisted technologies in the manuscript preparation process}

During the preparation of this work the author(s) used ChatGPT in order to improve the English language and readability. After using this tool/service, the author(s) reviewed and edited the content as needed and take(s) full responsibility for the content of the published article.

\bibliographystyle{unsrt}  
\bibliography{ref}  






\end{document}